\documentclass[conference]{IEEEtran}
\IEEEoverridecommandlockouts

\usepackage{cite}
\usepackage{amssymb,amsfonts}
\usepackage[tbtags]{amsmath}
\usepackage{graphicx}
\usepackage{textcomp}
\usepackage{xcolor}
\usepackage{amsthm}
\usepackage{bm}
\usepackage{graphicx}
\usepackage{float}
\usepackage{subfigure}
\usepackage{setspace} 
\usepackage{algorithm}
\usepackage{algorithmic}
\usepackage{bm}

\begin{document}

\title{Power Control and Random Serving Mode Allocation for CJT-NCJT Hybrid Mode Enabled Cell-Free Massive MIMO With Limited Fronthauls
}

\author{\IEEEauthorblockN{
Hangyu Zhang\IEEEauthorrefmark{1},
Rui Zhang\IEEEauthorrefmark{1}, 
Yongzhao Li\IEEEauthorrefmark{1},
Yuhan Ruan\IEEEauthorrefmark{1},
Tao Li \IEEEauthorrefmark{1},
and Dong Yang\IEEEauthorrefmark{1}
}
\IEEEauthorblockA{\IEEEauthorrefmark{1} School
	of Telecommunications Engineering, Xidian University, Xi'an, China \\
E-mail: hyzhang\_3@stu.xidian.edu.cn, \{rz, yzhli\}@xidian.edu.cn, ryh911228@163.com, taoli@xidian.edu.cn,\\ dyang@mail.xidian.edu.cn
\thanks{This work was supported in part by the Shaanxi Provincial Key Research and Development Programs (2023-ZDLGY-33, 2022ZDLGY05-03, 2022ZDLGY05-04 ) and the Youth Innovation Team of Shaanxi Universities. (\emph{Corresponding authors: Rui Zhang; Yongzhao Li.}).}
}}
\maketitle
\vspace{-10pt}
\begin{abstract}
With a great potential of improving the service fairness and quality for user equipments (UEs), cell-free massive multiple-input multiple-output (mMIMO) has been regarded as an emerging candidate for 6G network architectures. Under ideal assumptions, the coherent joint transmission (CJT) serving mode has been considered as an optimal option for cell-free mMIMO systems, since it can achieve coherent cooperation gain among the access points. However, when considering the limited fronthaul constraint in practice, the non-coherent joint transmission (NCJT) serving mode is likely to outperform CJT, since the former requires much lower fronthaul resources. In other words, the performance excellence and worseness of single serving mode (CJT or NCJT) depends on the fronthaul capacity, and any single transmission mode cannot perfectly adapt the  capacity limited fronthaul. To explore the performance potential of the cell-free mMIMO system with limited fronthauls by harnessing the merits of CJT and NCJT, we propose a CJT-NCJT hybrid serving mode framework, in which UEs are allocated to operate on CJT or NCJT serving mode. To improve the sum-rate of the system with low complexity, we first propose a probability-based random serving mode allocation scheme. With a given serving mode, a successive convex approximation-based power allocation algorithm is proposed to maximize the system's sum-rate. Simulation results demonstrate the superiority of the proposed scheme.

\end{abstract}

\begin{IEEEkeywords}
Cell-free massive MIMO, coherent joint transmission, non-coherent joint transmission, limited fronthaul, hybrid serving mode.
\end{IEEEkeywords}
\section{Introduction}
As an emerging paradigm for 6G network architectures, cell-free massive multiple-input multiple-output (mMIMO) can significantly improve user equipment (UE) service quality and fairness, in which multiple access points (APs) cooperate to provide service under the coordination management of the central processing unit (CPU)~\cite{cell-free,6G}.
The conception of cell-free mMIMO is initially proposed in~\cite{cell-free seminal}, wherein the coherent joint transmission (CJT) serving mode is adopted. In the CJT serving mode, since the APs that serving the certain UE transmit the same data and all the APs’ signal will naturally be coherently merged at the UE, the cell-free mMIMO system can benefit from the large coherent cooperation gain among APs.

Following this seminal work in~\cite{cell-free seminal}, researchers have focused on exploring the potential of the CJT serving mode in cell-free mMIMO systems. 
It has been validated that the channel hardening and favorable propagation characteristics observed in centralized mMIMO systems also hold in cell-free mMIMO systems, although the use of cell-free mMIMO slightly worsens the channel hardening~\cite{cell-free book}.
Accordingly, combined with specific objective power allocation algorithm, the conjugate beamforming (CB) precoding can perform well and is widely used in existing studies~\cite{APG,enhanced CB}. In~\cite{APG}, the accelerated projected gradient method was adopted to reduce the power allocation complexity considering four spectrum efficiency (SE)-related metrics. To further improve the SE, the authors in~\cite{enhanced CB} proposed an enhanced CB precoding, where the squared norm normalization is used to boost the channel hardening. On the whole, a great potential is shown in the cell-free mMIMO system under the CJT serving mode.

In addition to the studies considering ideal assumptions with CJT serving mode, the studies that consider the engineering constraints in practical deployment of cell-free mMIMO system also have attracted significant interest. One crucial constraint is that the fronthaul links (wired or wireless) between the CPU and APs have limited capacity~\cite{Limited Fronthaul,DL limited fronthaul1,DL limited fronthaul3}. Under limited wired fronthaul, the performance of CB precoding was analyzed in~\cite{DL limited fronthaul1} by using stochastic geometry.
The authors considered a hybrid microwave/mmWave wireless fronthaul in~\cite{DL limited fronthaul3}, demonstrating that the system deployment is mainly dominated by the APs with low fronthaul capacity.
However, in the CJT serving mode considered in~\cite{DL limited fronthaul1,DL limited fronthaul3}, the total fronthaul consumption of one specific UE equals to the product of this UE's data rate and the number of the serving APs, which will impose high requirements on the fronthaul capacity.

To further explore the potential of cell-free mMIMO system with limited fronthauls, another candidate serving mode, i.e., non-coherent joint transmission (NCJT), attracts attention recently~\cite{NCJT1,NCJT2}.
In the NCJT serving mode, since each AP that serves a particular UE transmit different part of the data~\cite{CJT outperforms NCJT1,CJT outperforms NCJT3}, the total fronthaul consumption of one specific UE equals to its data rate. Therefore, compared to CJT serving mode, NCJT serving mode has the advantage of reducing the fronthaul consumption. With the consideration of CB precoding, the authors in~\cite{NCJT and CJT compare2} compared the CJT serving mode and NCJT serving mode in terms of energy efficiency, showing that CJT is more sensitive to the fronthaul capacity. Furthermore, the authors in~\cite{NCJT and CJT compare1} optimized the sum-rate of the system in the NCJT serving mode and compared with that of the CJT serving mode, the results revel that NCJT can achieve considerable gains over CJT under limited fronthauls. The aforementioned works~\cite{NCJT and CJT compare2} and~\cite{NCJT and CJT compare1} establish a new perspective that NCJT can outperform CJT when considering limited fronthauls. However, the NCJT is capable of outperforming CJT only when the fronthaul capacity is relatively tight.

Based on the above literature review, we can conclude that the available fronthaul capacity determines the performance of CJT or NCJT serving mode. Whereas, existing works only support individual CJT or NCJT serving mode. Thus, how to harness both the advantages of the two serving mode simultaneously, i.e., the coherent cooperation gain of CJT serving mode and the low fronthaul consumption of NCJT serving mode, to further improve the performance of cell-free mMIMO systems with limited fronthauls merits further research. 

To fill this research gap, we propose a CJT-NCJT hybrid serving mode framework, where each UE is allowed to operate on either CJT serving mode or NCJT serving mode. Under the proposed framework, we first establish a signal model to characterize the multi-user interference arising from both CJT UEs and NCJT UEs. On this basis, we derive a general SE expression for the CJT-NCJT hybrid serving mode through hardening bound technique. To further improve the sum-rate of the system, fronthaul capacity-aware resource allocation scheme is also investigated in this paper. Specifically, we first propose a probability-based random serving mode allocation scheme to reduce the complexity. Then, a successive convex approximation (SCA)-based power allocation algorithm is designed to maximize the system's sum-rate with a given serving mode. It should be noted that the proposed random serving mode allocation can effectively exploit the fronthaul resource with extremely low overhead. With given system parameters, such as fronthaul capacity, the number of APs and UEs, we can pre-configure the probability of CJT serving mode to conduct mode allocation, which has wide practicability in terms of implementation.

\section{System Model}
We consider a cell-free mMIMO system comprised of $M$ APs equipped with $N$ antennas and $K$ single antenna UEs, wherein ${\cal{M}} = \left\{ {1, \ldots ,M} \right\}$ and ${\cal{K}} = \left\{ {1, \ldots ,K} \right\}$ denote the set of APs and UEs, respectively. To guarantee the scalability, we consider a UE-centric cell-free architecture, where each UE $k$ is served by a set of nearby APs denoted by ${\cal{M}}_{k} \subset {\cal{M}}$~\cite{UE centric cell-free1}. Then, the set of the UEs served by the $m$-th AP is represented as ${\cal{K}}_{m} \subset {\cal{K}}$.
As for the fronthaul, let $C_m^{\max }$ (bit/s/Hz) denote the limited fronthaul capacity from AP $m$ to CPU~\cite{NCJT and CJT compare2}. We consider the time-division duplex (TDD) protocol, where the coherence block is divided into three parts, which includes ${{\tau _{\rm{p}}}}$ for the channel estimation, ${{\tau _{\rm{u}}}}$ for the uplink (UL) transmission and ${{\tau _{\rm{d}}}}$ for the downlink (DL) transmission. The frequency-flat block fading channel model is adopted, where each coherence block has ${\tau _{\rm{c}}}$ symbols~\cite{cell-free book}.
The channel between the $m$-th AP and the $k$-th UE is denoted by ${{\bf{h}}_{m,k}} \sim \mathcal{CN} \left( {{{\bf{0}}_N},{{\bf{R}}_{m,k}}} \right)$ with ${{{\bf{R}}_{m,k}}}$ being the spatial covariance matrix. Then, the large-scale fading ${\beta _{m,k}}$ can be given by ${\beta _{m,k}} = \frac{1}{M}{\rm{tr}}\left( {{{\bf{R}}_{m,k}}} \right)$ with ${\rm{tr}}\left(  \cdot  \right)$ denotes the trace.
\vspace*{-0.15\baselineskip}
\subsection{Channel Estimation}
\vspace*{-0.15\baselineskip}
We assume that ${{\bm{\varphi}}_1},{{\bm{\varphi}}_2},...,{{\bm{\varphi}}_{\tau _{\rm{p}}}} \in {\mathbb{C}^{{{\tau _{\rm{p}}}} \times 1}}$ denote ${\tau _{\rm{p}}}$-length mutually orthogonal pilot sequences, each of which satisfies ${\left| {\bm{\varphi} _i} \right|^2} = {{\tau _{\rm{p}}}}$ with $i \in \left\{ {1,2, \ldots ,{\tau _p}} \right\}$. Then, let ${\bm{\varphi}_{{t_k}}}$ with ${t_k}\in \left\{ {1,2, \ldots ,{\tau _p}} \right\}$ denotes the index of the pilot sequence assigned to the $k$-th UE. Furthermore, it is assumed that ${\cal{P}}_k \subset \cal{K}$ stands for the set of UEs whose pilots are shared with the $k$-th UE, including itself. 
After all the UEs simultaneously transmit their pilots, the UL received pilot signal at AP $m$ can be expressed as
\begin{equation}
	{\bf{Y}}_m^{\rm{p}} = \sum\nolimits_{k = 1}^K {\sqrt {p_k^{\rm{p}}} } {{\bf{h}}_{m,k}}{\bm{\varphi }}_{{t_k}}^H + {\bf{N}}_m^{\rm{p}},
\end{equation}
where ${p_k^{\rm{p}}} \ge 0$ denotes the pilot power of the $k$-th UE and ${\bf{N}}_m^{\rm{p}} \in {\mathbb{C}^{N \times {{\tau _{\rm{p}}}}}}$ represents the received noise matrix with independent $\mathcal{CN}\left( {0,{\sigma _{{\rm{ul}}}^2}} \right)$ elements, where $\sigma _{{\rm{ul}}}^2$ represents the UL noise power. We adopt minimum mean square error (MMSE) channel estimation algorithm and the estimated channel can be given by~\cite{CJT outperforms NCJT1}
\begin{equation}
	\begin{split}
		{{{\bf{\hat h}}}_{m,k}} &= \sqrt {p_k^{\rm{p}}} {{\bf{R}}_{m,k}}{\left( {\sum\limits_{l \in {{\cal{P}}_k}} {{\tau _{\rm{p}}}p_l^{\rm{p}}{{\bf{R}}_{m,l}} + \sigma _{\rm{p}}^2} {{\bf{I}}_N}} \right)^{ - 1}}{\bf{Y}}_m^{\rm{p}}{\bm{\varphi}_{{t_k}}}   \\& \sim \mathcal{CN}\left( {{{\bf{0}}_N},{p_k^{\rm{p}}} {\tau _{\rm{p}}}{{\bf{R}}_{m,k}}{\bf{\Psi }}_{{t_k},m}^{ - 1}{{\bf{R}}_{m,k}}} \right),
		\vspace*{-0.55\baselineskip}  
	\end{split}
\end{equation}
where 
\begin{equation}
	{{\bf{\Psi }}_{{t_k},m}} = \mathbb{E} \left[ {{{{\bf{\hat y}}}_{m,{t_k}}}{\bf{\hat y}}_{{m,t_k}}^H}  \right] = \sum\nolimits_{l \in {{\cal{P}}_k}} {{\tau _{\rm{p}}}p_l^{\rm{p}}{{\bf{R}}_{m,l}}}  + \sigma _{\rm{ul}}^2 {\bf{I}},
\end{equation}
represents the correlation matrix of the ${{{\bf{\hat y}}}_{m,{t_k}}}$.

\vspace*{-0.15\baselineskip}
\subsection{Downlink Precoding}
\vspace*{-0.15\baselineskip}
Since we focus on the DL data transmission phase, it is reasonable to assume ${{\tau _{\rm{u}}}}=0$, i.e., ${{\tau _{\rm{d}}}} = {{\tau _{\rm{c}}}}-{{\tau _{\rm{p}}}}$. Due to the fully channel reciprocity in TDD, the estimated UL channels is used to calculate the DL precoding. To reduce the overhead of instantaneous CSI interaction from APs to CPU and the computational complexity of the precoding, CB precoding is employed, where the precoding is calculated at all APs locally. The precoding vector is given by 
\vspace*{-0.55\baselineskip}
\begin{equation}
	{{\bf{w}}_{m,k}}  = \begin{cases}
		{{{{{\bf{\hat h}}}_{m,k}}} \mathord{\left/{\vphantom {{{{{\bf{\hat h}}}_{m,k}}} {\sqrt {E\left[ {{{\left\| {{{{\bf{\hat h}}}_{m,k}}} \right\|}^2}} \right]} }}} \right.\kern-\nulldelimiterspace} {\sqrt {\mathbb{E}\left[ {{{\left\| {{{{\bf{\hat h}}}_{m,k}}} \right\|}^2}} \right]} }}  & m \in {{\cal M}_k}\\
		\bf{0}{\rm{ }}  & m \notin {{\cal M}_k} \end{cases},   
\end{equation}

\section{CJT-NCJT Hybrid Serving framework}
In this section, we first introduce the framework of the CJT-NCJT hybrid serving mode. Then, we establish a signal model for the hybrid serving mode. On this basis, we derive a general SE expression relying on hardening bound technique.
\vspace*{-0.15\baselineskip} 
\subsection{Hybrid Serving Framework}
\vspace*{-0.15\baselineskip}
Recall that under the CJT-NCJT hybrid serving mode, each UE is either operate on the CJT serving mode or the NCJT serving mode. Thus, all the UEs operating on the CJT serving mode can be considered as forming a group denoted by ${\cal{G}^{\rm{coh}}} \subset \cal{K}$. In a similar way, the group formed by all the UEs operating on the NCJT serving mode is denoted by ${{\cal{G}^{\rm{nc}}}} \subset \cal{K}$. Then, we have ${\cal{G}^{\rm{coh}}}\cup {{\cal{G}^{\rm{nc}}}} = {\cal{K}}$ and  ${\cal{G}^{\rm{coh}}} \cap {{\cal{G}^{\rm{nc}}}} = {\emptyset}$.

Next, we aim to analyze the overhead in the implementation of the CJT-NCJT hybrid serving mode. After all the APs transfer the statistical channel information to the CPU, the CPU determines the UEs' serving mode and the power coefficients. Then, the power coefficient and serving mode indication are delivered from the CPU. It should be noted that only an additional 1-bit information is required to notify each UE which serving mode to use, which is a very small burden on the system.

\vspace*{-0.15\baselineskip} 
\subsection{SE Analysis Under the CJT-NCJT Hybrid Serving Mode}
\vspace*{-0.15\baselineskip}
In this subsection we derive the generic SE expression of the hybrid CJT-NCJT serving mode. 
We first derive the SE expression for the UEs in the CJT serving mode group. For the UE $i \in {\cal{G}^{\rm{coh}}}$, the APs in the set ${\cal{M}}_i$ synchronously transmit the same data ${x_i}$ to the UE. For the $s$-th UE served by the NCJT serving mode, each AP $m \in {\cal{M}}_s$ transmits symbol ${{x_{m,s}}}$. Then, the received signal at UE $i$ with the CJT serving mode can be given by
\begin{equation}\label{CJT SE}	
\small	
	\begin{aligned}	
		y_i^{{\rm{coh}}} & \!=\!\!\! \sum\limits_{m \in {\cal M}_i^{}}\!\!\! {\sqrt {p_{m,i}^{}} {{ {{\bf{h}}_{m,i}^{H}} }}{\bf{w}}_{m,i}^{}x_i^{}} \! +\!\! \sum\limits_{i^{'} \in {{\cal G}^{{\rm{coh}}}}\hfill\atop
			i^{'} \ne i\hfill}\! {\sum\limits_{m \in {\cal M}_{i'}^{}}\!\!\! {\sqrt {p_{m,i'}^{}} {{ {{\bf{h}}_{m,i}^{H}} }}{\bf{w}}_{m,i'}^{}x_{i'}^{}} }  \\& + \sum\limits_{s \in {{\cal G}^{{\rm{nc}}}}} {\sum\limits_{m \in {\cal M}_s^{}}\!\!\! {\sqrt {p_{m,s}^{}} {{ {{\bf{h}}_{m,i}^{H}} }}{\bf{w}}_{m,s}^{}x_{m,s}^{}} }  + n_i^{},	
	\end{aligned}
\end{equation}
\normalsize
where \small $\mathbb{E}\left[{{{\left| {{x_i}} \right|}^2}} \right]=\mathbb{E}\left[{{{\left| {{x_{m,s}}} \right|}^2}} \right] = 1$\normalsize, and ${p_{m,i}}$ represents the power coefficients allocated by the $m$-th AP to UE $i$. A lower bound on the ergodic capacity, i.e., an achievable SE, can be calculated by using the hardening bounding technique~\cite{hardening bound,cell-free book}, where the large-scale fading is used to decode the signal. The hardening bound for the $i$-th CJT serving mode UE is given by $r_i^{{\rm{coh}}} = {{{\tau _d}} \mathord{\left/
		{\vphantom {{{\tau _d}} {{\tau _c}}}} \right.
		\kern-\nulldelimiterspace} {{\tau _c}}}{\log _2}\left( {1 + \gamma _i^{{\rm{coh}}}} \right)$, where ${\gamma _i^{\rm{coh}}}$ denotes the effective signal-interference-noise-ratio (SINR) of the $i$-th UE and can be expressed as
\vspace*{-0.15\baselineskip}
\begin{equation}\label{CJT SINR}
	\gamma _i^{{\rm{coh}}} = {{{{\left| {{\rm{D}}_i^{{\rm{coh}}}} \right|}^2}} \mathord{\left/
			{\vphantom {{{{\left| {{\rm{D}}_i^{{\rm{coh}}}} \right|}^2}} {\left( {{\rm{F}}_i^{{\rm{coh}}} - {{\left| {{\rm{D}}_i^{{\rm{coh}}}} \right|}^2} + {\rm{F}}_i^{{\rm{nc}}} + \sigma _i^2} \right)}}} \right.
			\kern-\nulldelimiterspace} {\left( {{\rm{F}}_i^{{\rm{coh}}} - {{\left| {{\rm{D}}_i^{{\rm{coh}}}} \right|}^2} + {\rm{F}}_i^{{\rm{nc}}} + \sigma _i^2} \right)}},
\vspace*{-0.15\baselineskip}
\end{equation} where
\vspace*{-0.15\baselineskip}
\begin{equation}
	{\rm{D}}_i^{{\rm{coh}}} = \sum\nolimits_{m \in {{\cal{M}}_i}} {\mathbb{E}\left[ {\sqrt {p_{m,i}^{}} {\bf{h}}_{m,i}^H{\bf{w}}_{m,i}^{}} \right]},
\vspace*{-0.15\baselineskip}
\end{equation} 
\vspace*{-0.15\baselineskip}
\begin{equation}
	{\rm{F}}_i^{{\rm{coh}}} = \sum\nolimits_{i' \in {{\cal G}^{{\rm{coh}}}}} {{\mathbb{E}}\left[ {{{\left| {\sum\nolimits_{m \in {\cal M}_{i'}^{}} {\sqrt {p_{m,i'}^{}} {{ {{\bf{h}}_{m,i}^{H}}}}{\bf{w}}_{m,i'}^{}} } \right|}^2}} \right]},
	\vspace*{-0.15\baselineskip}
\end{equation}
\vspace*{-0.15\baselineskip}
\begin{equation}
	{\rm{F}}_i^{{\rm{nc}}} = \sum\nolimits_{s' \in {{\cal G}^{{\rm{nc}}}}} {\sum\nolimits_{m \in {\cal M}_{s'}^{}} {{\mathbb{E}}\left[ {{{\left| {\sqrt {{p_{m,s'}}} {{{{\bf{h}}_{m,i}^{H}}}}{\bf{w}}_{m,s'}^{}} \right|}^2}} \right]}}.
	\vspace*{-0.15\baselineskip}
\end{equation}
In~(6), ${\left| {{\rm{D}}_i^{{\rm{coh}}}} \right|^2}$, ${\rm{F}}_i^{{\rm{coh}}}-{\rm{D}}_i^{\rm{coh}}$, ${\rm{F}}_i^{{\rm{nc}}}$ and ${\sigma _i^2}$ represent the signal gain, the sum of the interference from CJT UEs and the channel gain uncertain of the $i$-th UE, the interference from NCJT UEs, and noise power of the $i$-th UE, respectively.
\addtolength{\topmargin}{-0.03in}

Then, we derive the hardening bound for the NCJT serving mode UEs. The received signal of UE $s$ with the NCJT mode can be expressed as
\begin{equation}\label{NCJT SE}
\small	
\begin{aligned}
		y_{s}^{{\rm{nc}}} &\!\! =\!\!\!\!\! \sum\limits_{m \in {\cal M}_s^{}}\!\!\!\!\!\!{\sqrt {p_{m,s}^{}} {{ {{\bf{h}}_{m,s}^{H}} }}{\bf{w}}_{m,s}^{}x_{m,s}^{}} \! +\!\!\!\! \sum\limits_{s^{'} \in {{\cal G}^{{\rm{nc}}}}\hfill\atop s^{'} \ne s\hfill}\! {\sum\limits_{m \in {\cal M}_{s'}^{}}\!\!\!\!\!\! {\sqrt {p_{m,s'}^{}} {{ {{\bf{h}}_{m,s}^{H}} }}{\bf{w}}_{m,s'}^{}x_{m,s'}^{}} }  \\& + \sum\limits_{i \in {{\cal G}^{{\rm{coh}}}}} {\sum\limits_{m \in {\cal M}_i^{}}\!\!\!\!\! {\sqrt {p_{m,i}^{}} {{{{\bf{h}}_{m,s}^{H}} }}{\bf{w}}_{m,i}^{}x_i^{}} }  + n_s^{}.
	\end{aligned}
\end{equation}
\normalsize
Based on~\cite{NCJT and CJT compare2,hardening bound}, the hardening bound $r_{m,s}^{\rm{nc}}$ for $x_{m,s}$ can be given by $r_{m,s}^{{\rm{nc}}} = {{{\tau _d}} \mathord{\left/{\vphantom {{{\tau _d}} {{\tau _c}}}} \right.
\kern-\nulldelimiterspace} {{\tau _c}}}{\log _2}\left( {1 + \gamma _{m,s}^{{\rm{nc}}}} \right)$, where ${\gamma _{m,s}^{\rm{n}}}$ denotes the effective SINR and is given by
\small
\begin{equation}\label{NCJT SINR}
	\gamma _{m,s}^{{\rm{nc}}} = {{{{\left| {{\rm{D}}_{m,s}^{{\rm{nc}}}} \right|}^2}} \mathord{\left/
			{\vphantom {{{{\left| {{\rm{D}}_{m,s}^{{\rm{nc}}}} \right|}^2}} {\left( {{\rm{F}}_s^{{\rm{coh}}} + {\rm{F}}_s^{{\rm{nc}}} - \sum\nolimits_{n = 1}^m {{{\left| {{\rm{D}}_{n,s}^{{\rm{nc}}}} \right|}^2}}  + \sigma _s^2} \right)}}} \right.
			\kern-\nulldelimiterspace} {\left( {{\rm{F}}_s^{{\rm{coh}}} + {\rm{F}}_s^{{\rm{nc}}} - \sum\nolimits_{n = 1}^m {{{\left| {{\rm{D}}_{n,s}^{{\rm{nc}}}} \right|}^2}}  + \sigma _s^2} \right)}},		
\end{equation} \normalsize where \small
${\rm{D}}_{m,s}^{\rm{nc}} = \mathbb{E}\left[ {\sqrt {p_{m,s}^{}} {{ {{\bf{h}}_{m,s}^{H}} }}{\bf{w}}_{m,s}^{}} \right].$ \normalsize
Here, ${\rm{D}}_{m,s}^{\rm{nc}}$, ${{\rm{F}}_s^{{\rm{coh}}}}$, ${{\rm{F}}_s^{{\rm{nc}}} - \sum\nolimits_{n = 1}^m {{{\left| {{\rm{D}}_{n,s}^{{\rm{nc}}}} \right|}^2}} }$ and ${\sigma _s^2}$ stand for the signal gain of $s$-th UE, the interference from CJT UEs, the sum of the interference from NCJT UEs and the channel gain uncertain of the $s$-th UE, and noise power of the $s$-th UE, respectively. Thus, the total achievable SE of UE $s$ is given by $r_s^{{\rm{nc}}} = \sum\nolimits_{m \in {{\cal{M}}_s}} {r_{m,s}^{{\rm{nc}}}}$.

Based on~\cite{NCJT and CJT compare2}, the fronthaul consumption of AP $m$ in the CJT-NCJT hybrid serving mode can be calculated as 
\begin{equation}
	{C_m} = \sum\nolimits_{i \in {\cal{G}^{{\rm{coh}}}} \cap {{\cal{K}}_m}} {r_i^{{\rm{coh}}} + } \sum\nolimits_{s \in {\cal{G}^{{\rm{nc}}}} \cap {{\cal{K}}_m}} {r_{m,s}^{{\rm{nc}}}},
\end{equation}
which is the sum of the fronthaul consumption of the serving CJT UEs and NCJT UEs.

\section{Fronthaul Capacity-Aware Resource Allocation}
To explore the performance potential of the CJT-NCJT hybrid serving mode, we design a fronthaul capacity-aware resource allocation scheme in this section. First, we propose a random serving mode allocation scheme to reduce the complexity.
Then, a SCA-based power allocation scheme is designed to optimize the sum-rate of the cell-free mMIMO system with limited fronthauls under a given serving mode.

\subsection{Random Serving Mode Allocation}
Recall that the CJT and NCJT serving modes have the merits of coherent cooperation gain among APs and low fronthaul consumption, respectively. 
When the fronthaul capacity is sufficient or even unlimited, the system can obtain the maximum coherent cooperation gain when all UEs operate on the CJT serving mode. However, when the fronthaul capacity is limited, the constraint of the fronthaul capacity will limit the coherent cooperation gain of CJT serving mode, and the achievable SE of the system will decrease due to the great fronthaul consumption when all UEs executes CJT serving mode. In response to this situation, the advantage of NCJT's much lower fronthaul consumption can be exploited to improve the performance of the system by increasing the proportion of NCJT UEs in the hybrid serving mode. 
In other words, there exists a trade-off between coherent cooperation gain and the fronthaul capacity consumption. Specifically, for a given fronthaul capacity value, the proportion of UEs operate on CJT serving mode determines the coherent cooperation gain and the proportion of NCJT UEs mainly adapts the fronthaul consumption to satisfy the fronthaul constraint.

Since there exists mutual coupling between the serving mode and the power coefficients, the fronthaul capacity-aware resource allocation problem is a mixed-integer programming problem, which is challenging to be solved. For practical applications, the serving mode allocation scheme is expected to be implemented with low complexity and overhead. Based on the above analysis and from the perspective of the holistic performance of the system, we can divide the proportion of UEs executing CJT serving mode or NCJT serving mode to achieve the trade-off between the coherent cooperation gain and the fronthaul consumption. Therefore, we design a low complexity probability-based random serving mode allocation method, where the probability of a UE being allocated with the CJT serving mode is $p$, and the corresponding probability of this UE being allocated with the NCJT serving mode is $1-p$. According to the system parameters, such as the fronthaul capacity, we can determine the optimal probability $p$ by traversing over the values of the probability $p$. When $p$ is given, we can easily obtain the serving sets ${\cal{G}^{\rm{coh}}}$ and ${\cal{G}^{\rm{nc}}}$.

\vspace*{-0.15\baselineskip}
\subsection{Power Allocation with Given Serving Mode}
\vspace*{-0.15\baselineskip}
In this subsection, we aim to maximize the sum-rate of the system with a given serving mode by optimizing the power coefficients.
The sum-rate maximization problem can be formulated as
\vspace*{-0.35\baselineskip}
\begin{subequations} \label{Problem}
	\begin{align}
		\mathop {\max }\limits_{{\bf{P}} {\rm{ }}} \quad &\sum\limits_{i \in {\cal G}^{{\rm{coh}}}} {r_i^{{\rm{coh}}}}  + \sum\limits_{s \in {\cal G}^{{\rm{nc}}}} {\sum\limits_{m \in {{\cal M}_s}} {r_{m,s}^{{\rm{nc}}}} } \\
		{\rm{s}}{\rm{.t}}{\rm{.}} \quad 		
		&\sum\limits_{k \in {{\cal M}_m}} {p_{m,k}^{}}  \le P_m^{\max }{\rm{   }} \qquad  \forall m \in {\cal M},  \label{opti,c2} \\
		&{C_m} \le C_m^{\max }{\rm{   } } \quad \forall m \in {\cal M}, \label{opti,c3}, \\
		& {p_{m,k}} = 0,\forall k \in {\cal K}, \forall m \notin {{\cal M}_k},\label{opti,c4},
	\end{align}
\end{subequations}
in which matrix $\bf{P}$ collects all the power coefficients $p_{m,k}$ and  the optimal power coefficient can be obtained through solving~\eqref{Problem}. 
In~\eqref{Problem}, the constraints~\eqref{opti,c2}$\sim$\eqref{opti,c4} can be explained as follows:~\eqref{opti,c2} is the maximum power constraint of AP $m$; \eqref{opti,c3} ensures that the fronthaul consumption of AP $m$ is less than the capacity of the fronthaul; \eqref{opti,c4} is the power constraint arising from the considered UE-centric scenario.

\addtolength{\topmargin}{-0.03in}
To facilitate the optimization of the power coefficients, let us denote ${{\tilde p}_{m,k}} = \sqrt {{p_{m,k}}}$, and define ${{{\bf{\tilde p}}}_k} = {\left[ {{{\tilde p}_{1,k}}, \ldots ,{{\tilde p}_{M,k}}} \right]^T}$. Then the effective SINR of UE $i$ with the CJT serving mode can be rewritten as
\begin{equation}
	{\gamma}_i^{{\rm{coh}}}\!\! =\!\! \frac{{{{\left| {{\bf{\tilde p}}_i^T{\bf{b}}_i^{{\rm{coh}}}} \right|}^2}}}{{\sum\limits_{k \in {\cal G}^{{\rm{con}}}} {{\bf{\tilde p}}_k^T{\bf{C}}_{ki}^{{\rm{coh}}}} {{{\bf{\tilde p}}}_k} \!\!+ \!\!\sum\limits_{k \in {\cal G}^{{\rm{nc}}}} {{\bf{\tilde p}}_k^T{\rm{diag}}\left( {{\bf{c}}_{si}^{{\rm{nc}}}} \right)} {{{\bf{\tilde p}}}_k}\!\! +\!\! \sigma _i^2}},
\end{equation}
where ${{\bf{b}}_i^{{\rm{coh}}}}\in {\mathbb{R}^{{L \times 1}}}$, ${\bf{c}}_{si}^{{\rm{nc}}} \in {\mathbb{R}^{L \times 1}}$, and we have ${\left[ {{\bf{b}}_i^{{\rm{coh}}}} \right]_m} = 
		\mathbb{E}\left[ {{\bf{h}}_{m,i}^H{\bf{w}}_{m,i}^{}} \right]$ and
${\left[ {{\bf{c}}_{si}^{{\rm{nc}}}} \right]_m} = 
\mathbb{E}\left[ {{{\left| {{{ {{\bf{h}}_{m,i}^{H}} }}{\bf{w}}_{m,s}^{}} \right|}^2}} \right]$. Furthermore, when $k \ne i$, we define \small${\left[ {{\bf{C}}_{ki}^{{\rm{coh}}}} \right]_{l,r}}\!\! =\!\! 
		\mathbb{E}\left[ {{\bf{h}}_{l,i}^H{\bf{w}}_{l,k}^{}{\bf{w}}_{r,k}^H{\bf{h}}_{r,i}^{}} \right]$. \normalsize
Next, we define \small
${\left[ {{\bf{C}}_{ii}^{{\rm{coh}}}} \right]_{l,l}}\!\! =\!\!
		\mathbb{E}\left[ {{\bf{h}}_{l,i}^H{\bf{w}}_{l,k}^{}{\bf{w}}_{l,k}^H{\bf{h}}_{l,i}^{}} \right]\!\! - \!\! \left[ {{\bf{b}}_i^{{\rm{coh}}}} \right]_l^2$. \normalsize

Similarly, for the UE $s$ operating on NCJT serving mode, the effective SINR of ${x_{m,s}}$ can be rewritten as
\begin{equation}
	{\gamma}_{m,s}^{{\rm{nc}}}\! =\! \frac{{{{\left| {\tilde p_{m,s}^Tb_{m,s}^{{\rm{nc}}}} \right|}^2}}}{{\sum\limits_{k \in {\cal G}^{{\rm{con}}}} {{\bf{\tilde p}}_k^T{\bf{C}}_{ks}^{{\rm{coh}}}} {{{\bf{\tilde p}}}_k} \!+\! \sum\limits_{k \in {\cal G}^{{\rm{nc}}}} {{\bf{\tilde p}}_k^T{\rm{diag}}\left( {{\bf{C}}_{ks}^{{\rm{nc}}}\left( {:,m} \right)} \right)} {{{\bf{\tilde p}}}_k} \!+\! \sigma _s^2}},
\end{equation}
where $b_{m,s}^{{\rm{nc}}} \in \mathbb{R}$,  ${\bf{C}}_{ks}^{{\rm{nc}}} \in {\mathbb{R}^{L \times L}}$ with
$b_{m,s}^{{\rm{nc}}} = \mathbb{E}\left[ {{\bf{h}}_{m,s}^H{\bf{w}}_{m,s}^{}} \right]$,
\small
\begin{equation}
	\begin{split}
		&{\rm{ }}{\left[ {{\bf{C}}_{ks}^{{\rm{nc}}}\left( {:,m} \right)} \right]_n} \\
		&=\!\! \begin{cases}
			\mathbb{E}\left[ {{{\left| {{{ {{\bf{h}}_{n,s}^{H}} }}{\bf{w}}_{n,k}^{}} \right|}^2}} \right] - {\left| {b_{n,s}^{{\rm{nc}}}} \right|^2} & \!\!\! n \le m,m \in {{\cal M}_s},n \in {{\cal M}_s}\\
			\mathbb{E}\left[ {{{\left| {{{ {{\bf{h}}_{n,s}^{H}} }}{\bf{w}}_{n,k}^{}} \right|}^2}} \right] &\!\!\! n > m,m \in {{\cal M}_s}, n \in {{\cal M}_s}
		\end{cases}.
	\end{split}
\end{equation}
\normalsize

In problem~\eqref{Problem}, since the data rate expression in the objective function and constraints~\eqref{opti,c3} are non-convex, the first step to deal with this issue is to introduce auxiliary variables ${\left\{ {\mu _i^{{\rm{coh}}},\xi _i^{{\rm{coh}}},\vartheta _i^{{\rm{coh}}}} \right\}_{\left\{ {\forall i \in {\cal G}^{{\rm{con}}}} \right\}}}$,  ${\left\{ {\mu _{m,s}^{{\rm{nc}}},\xi _{m,s}^{{\rm{nc}}},\vartheta _{m,s}^{{\rm{nc}}}} \right\}_{\left\{ {\forall s \in {\cal G}^{{\rm{nc}}},\forall m \in {{\cal M}_s}} \right\}}}$ and then the problem~\eqref{Problem} can be transformed to
\vspace*{-0.35\baselineskip}
\begin{subequations} \label{Problem1_t}
	\begin{align}
		&\mathop {\max }\limits_{ {\left\{ {{{\tilde p}_{m,k}},\mu _i^{{\rm{coh}}},\xi _i^{{\rm{coh}}},\vartheta _i^{{\rm{coh}}}, }\right.} \hfill\atop { \left. \mu _{m,s}^{{\rm{nc}}},\xi _{m,s}^{{\rm{nc}}},\vartheta _{m,s}^{{\rm{nc}}} \right\}} }    \sum\limits_{i \in {\cal G}^{{\rm{con}}}} {\mu _i^{{\rm{coh}}}}  + \sum\limits_{s \in {\cal G}^{{\rm{nc}}}} {\sum\limits_{m \in {{\cal M}_s}} {\mu _{m,s}^{{\rm{nc}}}} } \label{opti_t} \\
		{\rm{s}}{\rm{.t}}{\rm{.}} & \frac{{{\tau _d}}}{{{\tau _c}}}{\log _2}\left( {1 + \xi _i^{{\rm{coh}}}} \right) \ge \mu _i^{{\rm{coh}}}{\rm{,  }} \;\forall i \in {\cal G}^{{\rm{coh}}},  \label{opti_t,c1}\\
		{\rm{ }}  &\frac{{{\tau _d}}}{{{\tau _c}}}{\log _2}\left( {1 + \xi _{m,s}^{{\rm{nc}}}} \right) \ge \mu _{m,s}^{{\rm{con}}}{\rm{,  }}\;\forall s \in {\cal G}^{{\rm{nc}}}{\rm{ }}, \;\forall m \in {{\cal M}_s}, \label{opti_t,c2} \\
		&{{{{\left| {{\bf{\tilde p}}_i^T{\bf{b}}_i^{{\rm{coh}}}} \right|}^2}} \mathord{\left/
				{\vphantom {{{{\left| {{\bf{\tilde p}}_i^T{\bf{b}}_i^{{\rm{coh}}}} \right|}^2}} {\vartheta _i^{{\rm{coh}}}}}} \right.
				\kern-\nulldelimiterspace} {\vartheta _i^{{\rm{coh}}}}} \ge \xi _i^{{\rm{coh}}},\forall i \in {\cal{G}^{{\rm{coh}}}}, \label{opti_t,c3}\\
		&{{{{\left| {{{\tilde p}_{m,s}}b_{m,s}^{{\rm{coh}}}} \right|}^2}} \mathord{\left/
				{\vphantom {{{{\left| {{{\tilde p}_{m,s}}b_{m,s}^{{\rm{coh}}}} \right|}^2}} {\vartheta _{m,s}^{{\rm{nc}}}}}} \right.
				\kern-\nulldelimiterspace} {\vartheta _{m,s}^{{\rm{nc}}}}} \ge \xi _{m,s}^{{\rm{nc}}},\;\forall s \in {\cal{G}^{{\rm{nc}}}},\;\forall m \in {{\cal{M}}_s}, \label{opti_t,c4}\\
		&{\vartheta _i^{{\rm{coh}}}} \ge \sum\limits_{k \in {\cal G}^{{\rm{con}}}} {{\bf{\tilde p}}_k^T{\bf{C}}_{ki}^{{\rm{coh}}}} {{{\bf{\tilde p}}}_k} + \sum\limits_{k \in {\cal G}^{{\rm{nc}}}} {{\bf{\tilde p}}_k^T{\rm{diag}}\left( {{\bf{c}}_{ki}^{{\rm{nc}}}} \right)} {{{\bf{\tilde p}}}_k} \nonumber\\ & \qquad + \sigma _i^2,\;\forall i \in {\cal G}^{{\rm{coh}}},  \label{opti_t,c5} \\
		&{\vartheta _{m,s}^{{\rm{nc}}}} \ge  {\sum\limits_{k \in {\cal G}_{}^{{\rm{nc}}}} {{\bf{\tilde p}}_k^T{\rm{diag}}\left( {{\bf{C}}_{ks}^{{\rm{nc}}}\left( {:,m} \right)} \right)} {{{\bf{\tilde p}}}_k}} + \sigma _s^2 \nonumber \\&
		\qquad + \sum\limits_{k \in {\cal G}^{{\rm{con}}}} {{\bf{\tilde p}}_k^T{\bf{C}}_{ks}^{{\rm{coh}}}} {{{\bf{\tilde p}}}_k},\; \forall s \in {\cal G}^{{\rm{nc}}}{\rm{ }},\;\forall m \in {{\cal M}_s}, \label{opti_t,c6} \\
		& \eqref{opti,c2}, \eqref{opti,c3}, \eqref{opti,c4}. \notag
	\end{align}
\end{subequations}
It should be emphasized that the equivalence between~\eqref{Problem1_t} and~\eqref{Problem} is due to the fact that constraints~\eqref{opti_t,c1}$\sim$\!~\eqref{opti_t,c6} hold with equality at the optimum point.
Among all the constraints,~\eqref{opti_t,c3} and~\eqref{opti_t,c4} are non-convex since the left hand side is a convex quadratic-over-linear function. To address this issue, we adopt the SCA technique to approximate the constraints as convex ones, which results in a high-performance solution with low complexity although
returning a local maximum solution. Specifically, let us first denote
\begin{equation}
	f_{}^{{\rm{coh}}}\left( {{{{\bf{\tilde p}}}_i},\vartheta _i^{{\rm{coh}}}} \right) = {{{{\left| {{\bf{\tilde p}}_i^T{\bf{b}}_i^{{\rm{coh}}}} \right|}^2}} \mathord{\left/
			{\vphantom {{{{\left| {{\bf{\tilde p}}_i^T{\bf{b}}_i^{{\rm{coh}}}} \right|}^2}} {\vartheta _i^{{\rm{coh}}}}}} \right.
			\kern-\nulldelimiterspace} {\vartheta _i^{{\rm{coh}}}}},
\end{equation}
and
\begin{equation}
	f_{}^{{\rm{nc}}}\left( {{{\tilde p}_{m,s}},\vartheta _{m,s}^{{\rm{nc}}}} \right) = {{{{\left| {{{\tilde p}_{m,s}}b_{m,s}^{{\rm{nc}}}} \right|}^2}} \mathord{\left/
			{\vphantom {{{{\left| {{{\tilde p}_{m,s}}b_{m,s}^{{\rm{nc}}}} \right|}^2}} {\vartheta _{m,s}^{{\rm{nc}}}}}} \right.
			\kern-\nulldelimiterspace} {\vartheta _{m,s}^{{\rm{nc}}}}}.
\end{equation}
Then, we approximate $f_{}^{{\rm{coh}}}\left( {{{{\bf{\tilde p}}}_i},\vartheta _i^{{\rm{coh}}}} \right)$ and  $f_{}^{{\rm{nc}}}\left( {{{\tilde p}_{m,s}},\vartheta _{m,s}^{{\rm{nc}}}} \right)$ by their first-order Taylor expansions to transform \eqref{opti_t,c3} and \eqref{opti_t,c4} to convex constraints. To this end, let ${\bf{\tilde p}}_i^{\left( t \right)},\vartheta _i^{{\rm{coh}},\left( t \right)},\tilde p_{m,s}^{\left( t \right)},\vartheta _{m,s}^{{\rm{nc}},\left( t \right)}$ be the values of ${{{\bf{\tilde p}}}_i},{\vartheta _i^{{\rm{coh}}}},{{\tilde p}_{m,s}},{\vartheta _{m,s}^{{\rm{nc}}}}$ at the $t$-th iteration of the SCA procedure. Then, we have two functions defined as follows:
\vspace*{-0.35\baselineskip} 
\small
\begin{equation}\label{SCA_1}
	\begin{split}
		&  F_{}^{{\rm{coh}}}\left( {{{{\bf{\tilde p}}}_i},{\vartheta _i^{{\rm{coh}}}};{\bf{\tilde p}}_i^{\left( t \right)},\vartheta_i^{{\rm{coh}},\left( t \right)}} \right) \\
		&  = f_{}^{{\rm{coh}}}\left( {{\bf{\tilde p}}_i^{\left(t \right)},\vartheta _i^{{\rm{coh}}, \left( t \right)}} \right) + \frac{{2{{\left( {{\bf{\tilde p}}_i^{\left( t \right)}} \right)}^T}{\bf{b}}_i^{{\rm{coh}}}{{\left( {{\bf{b}}_i^{{\rm{coh}}}} \right)}^T}}}{{\vartheta _i^{{\rm{coh}},\left( t \right)}}}\left( {{{{\bf{\tilde p}}}_i} - {\bf{\tilde p}}_i^{\left( t \right)}} \right) \\ 
		& \qquad - {{f_{}^{{\rm{coh}}}\left( {{\bf{\tilde p}}_i^{\left( t \right)},\vartheta _i^{{\rm{coh}},\left( t \right)}} \right)\left( {\vartheta _i^{{\rm{coh}}} - \vartheta _i^{\left( t \right)}} \right)} \mathord{\left/
				{\vphantom {{f_{}^{{\rm{coh}}}\left( {{\bf{\tilde p}}_i^{\left( t \right)},\vartheta _i^{{\rm{coh}},\left( t \right)}} \right)\left( {\vartheta _i^{{\rm{coh}}} - \vartheta _i^{\left( t \right)}} \right)} {\vartheta _i^{{\rm{coh}},\left( t \right)}}}} \right.
				\kern-\nulldelimiterspace} {\vartheta _i^{{\rm{coh}},\left( t \right)}}},
	\end{split}
\normalsize
\vspace{-10pt}
\end{equation}
and
\vspace*{-0.55\baselineskip} 
\begin{equation}\label{SCA_2}
	\small
	\begin{split}
		& F_{}^{{\rm{nc}}}\left( {{{\tilde p}_{m,s}},{\vartheta _{m,s}^{{\rm{nc}}}};\tilde p_{m,s}^{\left( t \right)},\vartheta _{m,s}^{{\rm{nc}},\left( t \right)}} \right)\\
		&= f_{}^{{\rm{nc}}}\left( {\tilde p_{m,s}^{\left( t \right)},\vartheta _{m,s}^{{\rm{nc}},\left( t \right)}} \right) + \frac{{2b_{m,s}^{{\rm{nc}}}{{\tilde p}_{m,s}}}}{{\vartheta _{m,s}^{{\rm{nc}},\left( t \right)}}}\left( {{{\tilde p}_{m,s}} - \tilde p_{m,s}^{\left( t \right)}} \right) \\
		& \qquad - {{f_{}^{{\rm{nc}}}\left( {\tilde p_{m,s}^{\left( t \right)},\vartheta _{m,s}^{\left( t \right)}} \right)\left( {\vartheta _{m,s}^{{\rm{nc}}} - \vartheta _{m,s}^{\left( t \right)}} \right)} \mathord{\left/
				{\vphantom {{f_{}^{{\rm{nc}}}\left( {\tilde p_{m,s}^{\left( t \right)},\vartheta _{m,s}^{\left( t \right)}} \right)\left( {\vartheta _{m,s}^{{\rm{nc}}} - \vartheta _{m,s}^{\left( t \right)}} \right)} {\vartheta _{m,s}^{{\rm{nc}},\left( t \right)}}}} \right.
				\kern-\nulldelimiterspace} {\vartheta _{m,s}^{{\rm{nc}},\left( t \right)}}}.
	\end{split}
\vspace*{-0.55\baselineskip}
\end{equation}
\normalsize
Then, let us replace the right hand side of~\eqref{opti_t,c3} and~\eqref{opti_t,c4} with~\eqref{SCA_1} and~\eqref{SCA_2}, respectively. We can obtain
\begin{equation} \label{TC1}
	F_{}^{{\rm{coh}}}\left( {{{{\bf{\tilde p}}}_i},{\vartheta _i^{{\rm{coh}}}};{\bf{\tilde p}}_i^{\left( t \right)},\vartheta _i^{{\rm{coh}},\left( t \right)}} \right) \ge \xi _i^{{\rm{coh}}},\forall i \in {\cal G}_t^{{\rm{coh}}},
\end{equation}
and
\begin{equation} \label{TC2}
	F_{}^{{\rm{nc}}}\left( {{{\tilde p}_{m,s}},{\vartheta _{m,s}^{{\rm{nc}}}};\tilde p_{m,s}^{\left( t \right)},\vartheta _{m,s}^{{\rm{nc}}, \left(t\right)}} \right) \ge \xi _{m,s}^{{\rm{nc}}},\forall s \in {\cal G}_t^{{\rm{nc}}}{\rm{ }}, \forall m \in {{\cal M}_s},
\end{equation}
which are convex constraints. Therefore, the problem~\eqref{Problem1_t} is transformed to a convex optimization problem by using~\eqref{TC1},~\eqref{TC2} instead of~\eqref{opti_t,c3} and~\eqref{opti_t,c4} at the $t$-th iteration of the SCA algorithm. Based on the above reformulations, the optimization problem to be solved at the $t$-th iteration can be given by
\vspace*{-0.55\baselineskip}
\begin{subequations} \label{nth SCA}
	\begin{align} 
		&\mathop {\max }\limits_{ {\left\{ {{{\tilde p}_{m,k}},\mu _i^{{\rm{coh}}},\xi _i^{{\rm{coh}}},\vartheta _i^{{\rm{coh}}}, }\right.} \hfill\atop { \left. \mu _{m,s}^{{\rm{nc}}},\xi _{m,s}^{{\rm{nc}}},\vartheta _{m,s}^{{\rm{nc}}} \right\}}  }   \sum\limits_{i \in {\cal G}^{{\rm{con}}}} {\mu _i^{{\rm{coh}}}}  + \sum\limits_{s \in {\cal G}^{{\rm{nc}}}} {\sum\limits_{m \in {{\cal M}_s}} {\mu _{m,s}^{{\rm{nc}}}} } \\
		&{\rm{s}}{\rm{.t}}{\rm{.}}   \; \eqref{opti_t,c1}, \eqref{opti_t,c2}, \eqref{TC1}, \eqref{TC2}, \eqref{opti_t,c5}, \eqref{opti_t,c6}, \eqref{opti,c2} \sim \eqref{opti,c4},
	\end{align}
\end{subequations}
which is a convex optimization problem. It can be efficiently solved by using the interior point method, which can be easily implemented through the existing optimization tools such as CVX~\cite{CVX}. 
In summary, the proposed SCA-based power allocation algorithm is shown in Algorithm 1.
\vspace{-7pt}
\begin{algorithm}[h]
	\caption{The SCA-based power allocation algorithm}
	\begin{algorithmic}[1]
		\renewcommand{\algorithmicrequire}{\textbf{Input:}}
		\REQUIRE { ${{\cal G}^{{\rm{coh}}}}$, ${{\cal G}^{{\rm{nc}}}}$}.\\
		\renewcommand{\algorithmicrequire}{\textbf{Initialize:}}
		\REQUIRE {Set $t=0$, ${{{\bf{\tilde P}}}^{\left( 0 \right)}} = zeros\left( {K,M} \right)$, and calculate ${\bf{\tilde p}}_i^{\left( 0\right)}$, ${\vartheta_i^{{\rm{coh}}, \left( 0 \right)}}$, $\tilde p_{m,s}^{\left( 0 \right)}$, $\vartheta _{m,s}^{{\rm{nc}}, \left( 0 \right)}$, $\xi _i^{{\rm{coh,}}\left( {0} \right)}$, $\xi _{m,s}^{{\rm{nc,}}\left( {0} \right)}$.}
		\REPEAT
		\STATE $t \leftarrow t + 1$;
		\STATE Solve the convex optimization problem~\eqref{nth SCA};
		\STATE Update ${\bf{\tilde p}}_i^{\left( t \right)}$, ${\vartheta_i^{{\rm{coh}}, \left( t \right)}}$, $\tilde p_{m,s}^{\left( t \right)}$, $\vartheta _{m,s}^{{\rm{nc}}, 
			\left( t \right)}$, $\xi _i^{{\rm{coh,}}\left( {t} \right)}$, $\xi _{m,s}^{{\rm{nc,}}\left( {t} \right)}$;
		\UNTIL Convergence of~\eqref{nth SCA} or reach the maximum number of iterations ${T^{\max }}$.
		\renewcommand{\algorithmicrequire}{\textbf{Output:}}
		\REQUIRE { Power coefficients ${{{\bf{ P}}}^\star }\left( {{\cal G}^{{\rm{coh}}}}, {{\cal G}^{{\rm{nc}}}} \right)$}.	
	\end{algorithmic}
\end{algorithm}
\vspace{-8pt}
\subsection{Convergence and Complexity}
\vspace{-3pt}
Since~\eqref{TC1},~\eqref{TC2} satisfy the conditions in~\cite[Section III-C, Assumptions 1-3]{SCA convergence 1}, it is guaranteed that the $t$-th SCA process to solve~\eqref{Problem1_t} can convergence.
By using the interior point method, the complexity of Algorithm 1 is given by \small $  \mathcal{O}\left({T^{\max }} {\log \left( {{1 \mathord{\left/{\vphantom {1 \varepsilon }} \right.\kern-\nulldelimiterspace} \varepsilon }} \right){{\left( {\sum\limits_{k \in {\cal K}} {\left| {{{\cal M}_k}} \right|}  + 3\left| {{{\cal G}^{{\rm{con}}}}} \right| + 3\sum\limits_{s \in {{\cal G}^{{\rm{nc}}}}} {\left| {{{\cal M}_s}} \right|} } \right)}^{3.5}}} \right)$ \normalsize  with $\varepsilon$ controls the accuracy of the solution~\cite{CVX}.
\vspace{-3pt}
\section{Simulation results}
\vspace*{-0.3\baselineskip}
In this section, we present simulation results to demonstrate the performance of the proposed CJT-NCJT hybrid serving mode along with the resource allocation scheme. A cell-free mMIMO system where APs and UEs are uniformed distributed in a 600~m$\times$600~m square is modeled. We use wrap-around technique to eliminate the bound effect~\cite{NCJT and CJT compare2}. The set of APs that serve each UE $k$ is formed by selecting $\left| {{\cal{M}}_k}\right|$ APs with the largest large-scale fading value $\beta_{m,k}$. The coherence blocks have ${\tau _c}=200$ samples, and the number of orthogonal pilots is set as ${\tau _p}=10$. In terms of pilot assignment, random ${\tau _p}$ UEs have orthogonal pilots and the remaining UEs are random. The pilot power and the maximum AP transmit power are~$p_k^{\rm{p}}=$0.1W and~$P_m^{\rm{max}}=$0.2W, respectively. Without loss of generality, we assume that the fronthaul capacity between CPU and all the APs are equal, i.e., $C_m^{{\rm{max}}} = {C^{\rm{max}}}$.

For the channel model, the large-scale coefficients are modeled according to~\cite[Eq (2.16)]{cell-free book}.
For the spatially correlated, we use Gaussian local scattering model with 15$^\circ $ standard derivation to calculate the channel covariance matrix. In addition, the bandwidth is $B=20$~MHz and the noise power is given by ${\sigma ^2} = B{\kappa _{\rm{B}}}{T_0}{\sigma _F}$, with ${\kappa _{\rm{B}}}=1.381\times{10^{-23}}$~Joule per Kelvin, ${T_0} = 290$~Kelvin and ${\sigma _F}=$9~dB.

\begin{figure}
	\vspace{-5pt}
	\centering
	\includegraphics[width=0.42\textwidth]{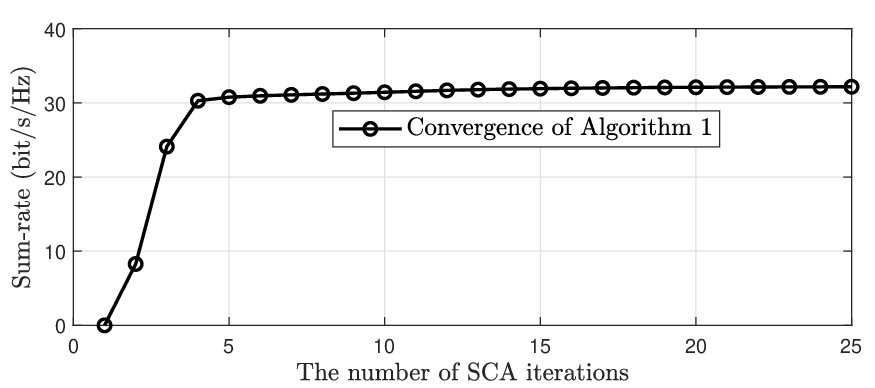}
	\vspace{-10pt}
	\caption{The convergence of the Algorithm 1 when $M=14$, $N=8$, $K=15$, $|\mathcal{M}_k|=8$ and ${C^{\rm{max}}} = 15$.}
  \vspace{-15pt}
\end{figure}
First, the convergence of the SCA-based power allocation algorithm (Algorithm 1) is verified in Fig.~1, where the serving mode is set as $[010101010101010]$ with '0' representing NCJT and '1' standing for the CJT. As expected, Algorithm 1 rapidly converges to a stable solution after several of iterations.

\addtolength{\topmargin}{0.13in}
\begin{figure}
\vspace{-7pt}
  \centering
  \includegraphics[width=0.42\textwidth]{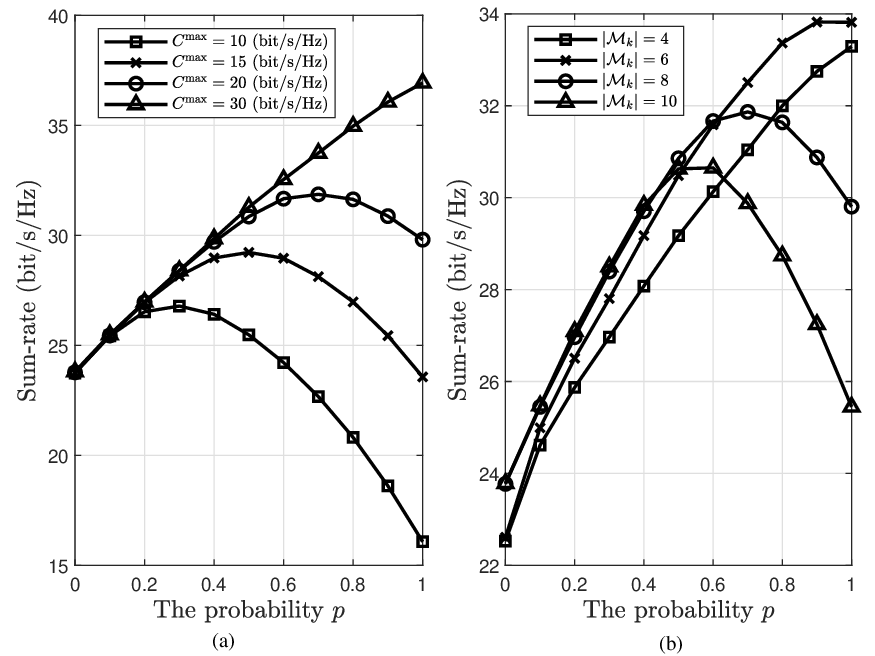}
  \vspace{-10pt}
  \caption{The sum-rate of the proposed CJT-NCJT hybrid serving mode with random serving mode allocation versus the the probability $p$, in which (a): when varying ${C^{\rm{max}}}$ with $M=14$, $N=8$, $K=15$ and $|\mathcal{M}_k|=8$; (b): when varying $|\mathcal{M}_k|$ with $M=14$, $N=8$, $K=15$ and ${C^{\rm{max}}}=20$~bit/s/Hz.}
\vspace{-15pt}
\end{figure}

Then, the optimum probability $p$ is illustrated under different system parameters including fronthaul capacity ${C^{\rm{max}}}$ and the number of serving APs $|\mathcal{M}_k|$ in Fig.~2. It should be noted that $p=1$ means that all the UEs execute CJT serving mode and $p=0$ stands for that all the UEs operate on NCJT serving mode.
From Fig.~2(a), it can be seen that when ${C^{\rm{max}}}$ is sufficient (${C^{\rm{max}}}=30$~bit/s/Hz), the sum-rate monotonically increases with the increase of $p$ until all the UEs execute CJT serving mode. The reason is that sufficient fronthaul capacity can support the system to obtain the maximum coherent cooperation gain.
Furthermore, it can be observed that the probability $p$ corresponding to the maximum sum-rate of the system decreases with the decrease of ${C^{\rm{max}}}$. The reason is that the tighter the fronthaul capacity, the performance of CJT will also be limited. To improve the sum-rate, increasing the proportion of NCJT UEs can benefit more UEs.
Meanwhile, we can find that, with the decrease of ${C^{\rm{max}}}$, the 
maximum sum-rate of the system decreases. This is because that the system's performance is limited by the fronthaul capacity.

From Fig.~2(b), it can be observed that when the number of serving APs is small ($|\mathcal{M}_k|=4$), the sum-rate monotonically increase with the increase of $p$ until all UEs execute CJT. This is because that, when $|\mathcal{M}_k|$ is small, the number of UEs to be served by each AP $\mathcal{K}_m$ also is small, and the fronthaul capacity is sufficient to support all the served UEs to execute CJT. Furthermore, we can find that the probability $p$ corresponding to the maximum sum-rate of the system decreases with the increase of $\mathcal{M}_k$. This can be explained as follows. As $\mathcal{M}_k$ increases, $\mathcal{K}_m$ also increases. At this point, the fronthaul capacity of each AP needs to support more UEs. Thus, compared with the coherent cooperation gain of the CJT, the low fronthaul consumption advantage of the NCJT serving mode can benefit more UEs. Finally, it also can be found that with the increase of $\mathcal{M}_k$, the maximum sum-rate first increases and then decreases. This is because that the sum-rate of the hybrid serving mode is composed of the sum-rate of the CJT UEs and that of the NCJT UEs. The SE of the NCJT UEs can be guaranteed in a relative low fronthaul capacity regime due to the low fronthaul consumption advantage. However, when considering the CJT UEs, the limitation of the fronthaul capacity causes the SE of each CJT UE decrease with the increase of $\mathcal{K}_m$ due to the high fronthaul consumption. At this point, the coherent cooperation gain will tend to reduce. Combining the situation of CJT and NCJT UEs, there exists an optimal number on $\mathcal{M}_k$ that enables both CJT and NCJT to operate at their maximum potential simultaneously in the hybrid mode.

\section{Conclusion}
This work has proposed a CJT-NCJT hybrid serving framework that combines the advantages of the CJT and NCJT serving modes for cell-free mMIMO systems with limited fronthauls. To explore the performance potential of the hybrid serving mode, we have proposed a random serving mode allocation scheme along with a SCA-based power allocation algorithm. The simulation results have demonstrated that even with the simplest random serving mode allocation scheme, the hybrid serving mode can show superior performance in term of system's sum-rate.

\end{document}